\documentclass[useAMS,usenatbib]{mn2e}
\usepackage{epsfig}
\usepackage{graphicx}
\usepackage{epstopdf}

\def\CI{{\it CI }}
\def\CD{{\it CD }}
\def\WD{{\it WD }}
\def\CH{{\it CH }}
\def\CCH{{\it CCH }}
\def\vsigma{V_{rot}/\sigma}
\newcommand{\ltorder}{\hbox{ \rlap{\raise 0.425ex\hbox{$<$}}\lower
0.65ex\hbox{$\sim$} }}
\newcommand{\gtorder}{\hbox{ \rlap{\raise 0.425ex\hbox{$>$}}\lower
0.65ex\hbox{$\sim$} }}

\newcommand\apj{{ApJ}}%   
\newcommand\aap{{A\&A}}%   
\newcommand\mnras{{MNRAS}}%   
\title[Early star cluster dynamical evolution]{Kinematical fingerprints of star cluster early dynamical evolution}
\author[E.Vesperini et al.] {Enrico Vesperini$^1$, Anna Lisa Varri$^{1,2}$, Stephen L.W. McMillan$^3$, Stephen E. Zepf$^4$\\
$^1$ Department of Astronomy, Indiana University, Bloomington, IN 47405, USA\\
$^2$ School of Mathematics and Maxwell Institute for Mathematical Sciences, University of Edinburgh, Edinburgh EH9 3JZ, UK\\
$^3$Department of Physics, Drexel University, Philadelphia, PA 19104, USA\\
$^{4}$Department of Physics and Astronomy, Michigan State University, East Lansing, MI 48824, USA\\}
\begin{document}
\maketitle

\begin{abstract}
We study the effects of the external tidal field on the violent relaxation phase of star clusters dynamical evolution, with 
particular attention to the kinematical properties of the equilibrium
configurations emerging at the end of this  phase. 
We show that star clusters undergoing the
process of violent relaxation in the tidal field of 
their host galaxy can acquire significant internal differential rotation and are
characterized by a distinctive radial variation of the velocity 
anisotropy. 
These kinematical properties are the result  
of the symmetry breaking introduced by the external tidal field in the
collapse phase and of the action of the 
Coriolis force on the orbit of the stars.  
The resulting equilibrium configurations
are characterized by differential rotation, with a peak located
between one and two half-mass radii.
As for the anisotropy, similar to clusters evolving in
isolation, the systems explored in this Letter are characterized by an
inner isotropic core, followed by a region of increasing radial
anisotropy. However for systems evolving in an external tidal field
the degree of radial anisotropy reaches a maximum in the
cluster intermediate regions and then progressively decreases, with the cluster outermost
regions being characterized by isotropy or a mild tangential anisotropy.    
Young or old but less-relaxed dynamically young star clusters may keep memory of these
kinematical fingerprints of their early  dynamical evolution. 
\end{abstract}
\begin{keywords}
globular clusters: general 
\end{keywords}
\section{Introduction}
Following the pioneering work of Lynden-Bell (1967) on the fundamental
aspects of the evolution of stellar systems toward a virial
equilibrium state, this phase of early 
relaxation of stellar systems (often referred to as violent
relaxation) has been intensively investigated for systems spanning
a broad range of physical scales  and environments. 
A large number of studies have explored this process in the context of
galaxy formation and early evolution (see e.g. Bertin 2000, Binney \&
Tremaine 2008 and references therein for reviews of the vast
literature devoted to this topic).  

Other studies have focussed their attention on star clusters (see
e.g. Aarseth et al. 1988, Boily et al. 1999) and addressed a number of additional
questions relevant to these smaller stellar systems. Examples include the
possible connection  between this dynamical phase and early mass
segregation (see e.g. McMillan et al. 2007, Moeckel \& Bonnell, 2009, Allison et al. 2009), the evolution of the number and orbital properties of
primordial binary stars (Vesperini \& Chernoff 1996, Kroupa et
al. 1999, Kroupa \& Burkert 2001, Parker et al. 2011), the dependence of the depth and dynamics of the initial collapse on the number of particles (Aarseth et al. 1988),
the frequency and implications of stellar collisions (Fujii \& Portegies Zwart 2013).

In studies of star cluster violent relaxation,
the effects of the star cluster's host galaxy tidal field are rarely considered (Theis 2002; see also Boily \& Pichon 2001 for a study of the effect of the host galaxy's torque on a triaxial star cluster). 
The goal of this Letter is to show how  the dynamics of a star cluster
during the violent relaxation phase and the kinematical properties of the
equilibrium system emerging at the end of this process
are affected by the host galaxy tidal field. 

 We will show that a cluster can acquire a significant differential
rotation during the violent relaxation phase. 
In addition we will show that the velocity distribution anisotropy in isolated clusters and in clusters affected by the host galaxy tidal field differ significantly.
Isolated
stellar systems emerge from the violent relaxation phase with
a velocity distribution increasingly  radially anisotropic at larger
distances from the center (see e.g. van Albada 1982, Trenti et al. 2005);
in clusters affected by an 
external tidal field, the velocity anisotropy at the end of the violent relaxation phase is qualitatively similar to that
of isolated clusters in the 
innermost (isotropic) and intermediate (radially anisotropic) regions
but, in contrast with isolated clusters, the outer regions are characterized by an isotropic or a slightly
tangentially anisotropic velocity distribution.

The outline of this Letter is the following. In Section 2 we describe
the method and the initial conditions adopted in our study. In Section 3 we
present our results. Section 4 summarizes our conclusions.
\section {Method and Initial Conditions}
For our investigation we have carried out a number of N-body simulations using
the {\tt starlab} package (Portegies Zwart et al. 2001) accelerated by a
GPU (Gaburov et al. 2009).

The systems considered in this study are assumed to move on  circular
orbits in an external  tidal field equal to either that of an isothermal halo or that of the  local Galactic potential
including the potential of the Galactic disk (assuming, in the latter
case, an orbit on the plane of the Galactic disk at a distance from
the Galactic center equal to that of the Sun).

In a coordinate system centered
on the center of mass of the cluster and rotating around the Galactic
center with constant angular velocity $\omega$, with the $x$ and $y$ axes
pointing away from the Galactic center and in the direction of motion
respectively,  the equations of motion for a star in the cluster,
 can be
written in the tidal approximation as (see e.g. Heggie \& Hut 2003)

\begin{eqnarray}
\ddot{x_i}& = & F_x-\alpha_1x_i+2\omega\dot{y_i} \\
\ddot{y_i}& = & F_y-2\omega\dot{x_i} \\
\ddot{z_i}& = & F_z-\alpha_3 z_i
\end{eqnarray}

where $(F_x,~F_y,~F_z)$ denote the force due to the other cluster stars, $\alpha_1=-2 \omega^2$  for an isothermal halo and
$\alpha_1=-4A(A-B)$ for the local Galactic disk field (where $A$ and
$B$ are the Oort constants; see e.g. Heggie \& Hut 2003), and
$\alpha_3=-0.5 \alpha_1$ for the isothermal halo and $\alpha_3=4\pi G
\rho_D+2(A^2-B^2)$ for the local Galactic disk (where $\rho_D$ is the
local density). All kinematical properties discussed in this Letter for clusters evolving in a tidal field, including those of initial conditions, are calculated in the rotating coordinate system introduced above (unless otherwise specified) and Eqs. 1-3  will be used in discussing the results of our simulations.
In describing the rotational properties of the clusters, we will refer
to prograde (retrograde) rotation to
indicate rotation in the same sense as (opposite sense to) that of the
cluster around the Galactic center (the rotational
velocities calculated in the adopted rotating 
coordinate system are relative
to the cluster initial synchronous solid-body rotation with angular velocity
$\omega$).

In addition to the cluster density profile, we characterize the
initial conditions explored by using the virial ratio, $Q=T/\vert V \vert$ (where $T$ is the total
kinetic energy and $V$ is the cluster internal potential energy), and the
ratio of the cluster initial limiting radius to the Jacobi radius, $R_L/R_J$.

We have carried out four simulations starting with an equal-mass,
spherical, homogeneous system, with a total number of particles
$N=60,000$, $R_L/R_J=0.5$ and the following combinations of $(Q,
\hbox{Tidal-Field})$: 
$(0.01,\hbox{Disk})$, $(0.1,\hbox{Disk})$, $(0.01,\hbox{Halo})$,
$(0.01,\hbox{Isolated})$. Hereafter, we will 
refer to these simulations as, respectively, \CD (Cold-Disk), \WD
(Warm-Disk), \CH 
(Cold-Halo), \CI (Cold-Isolated). The \CI simulation has been included
as a reference case to better illustrate the differences
between the results obtained with and without an external tidal
field. 
An additional simulation with an initial clumpy density distribution,
$R_L/R_J=0.5$ and $(Q, \hbox{Tidal-Field})=(0.01,\hbox{Halo})$,
hereafter referred to as \CCH (Cold-Clumpy-Halo) has been run. The
clumpy density distribution has been produced simply by letting the initial
Poisson density fluctuations in an isolated homogeneous sphere grow
until a time approximately equal to $0.97 t_{ff}$ (where $t_{ff}=\sqrt{3\pi/(32G\rho)}$ is the free-fall timescale) and then resetting the virial ratio to the desired value for the 
initial conditions used for the simulation.  

The units adopted are such that the
total mass of the system, $M$, the gravitational constant $G$ are equal to 1 and the initial limiting radius $R_L$ is equal to about 1.2 . In these units the
initial free-fall timescale is equal to about 1.45.

Particles moving beyond a radius
equal to twice the Jacobi radius are removed from the simulation.

We point out that this Letter is aimed at discussing
only the fundamental aspects of a star cluster
violent relaxation in an external tidal field. 
A complete survey of simulations fully exploring the parameter space of initial structural and kinematical properties is
currently in progress and will be presented in a separate paper. 
\section{Results}
Inspection of the equations of motions (eqs.1-3) discussed in the
previous section, 
shows that the presence of an external tidal field and of the associated initial synchronous rotation adopted alters  the
collapse of a cold system in two ways. First, the cluster
collapse is not isotropic:  the collapse 
is accelerated in the $z$ direction and slowed down in the $x$
direction. Second, the Coriolis force affects the motion in the $x$
and $y$ directions by deflecting the star radial motion in prograde rotation during the collapse and in the 
opposite direction when the systems re-expands.

\begin{figure}    
\centering{
\includegraphics[width=8.8cm]{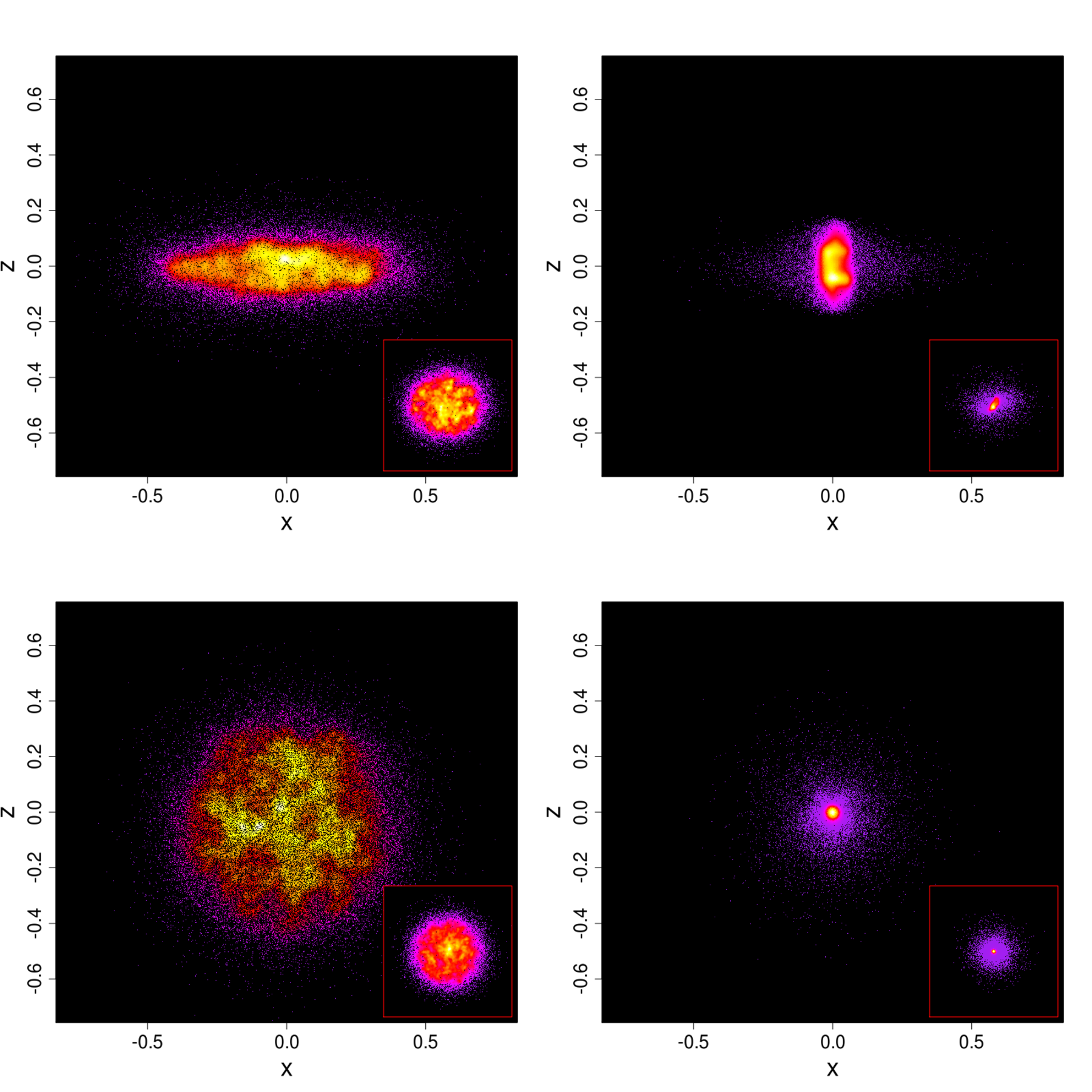}
}
\caption{Snapshots of the simulations \CD (top panels) and \CI (bottom
  panels) at two times ($t\simeq 1.31$ left panels; $t\simeq 1.47$
  right panels) close to the maximum contraction phase of the
  collapse. Particles are colored according to the local projected
  density (the density increases from purple to yellow) and in each panel the range of colors represents the range
  of densities at that time. 
Main figure in each panel shows the $x-z$
  projection; each inset shows the $x-y$ projection.} 
\label{fig:snapsh}
\end{figure}

The first effect is shown in Fig.\ref{fig:snapsh}.
This figure illustrates the difference between the collapse of
an isolated system ({\it CI}; lower panels) and that of a system affected by an external tidal
field ({\it CD}; upper panels) by showing two snapshots at times around the
moment of  maximum contraction. While the \CI system maintains its initial spherical symmetry, the \CD system flattens in the
$z$ direction first (top left panel); as the re-expansion phase in the $z$
direction starts the \CD system is still collapsing in the $x-y$
directions (top right panel) producing a significant elongation in the $z$
direction. In each panel, the inset shows the system projection on the $x-y$ plane.   

\begin{figure}    
\centering{
\includegraphics[width=8.8cm]{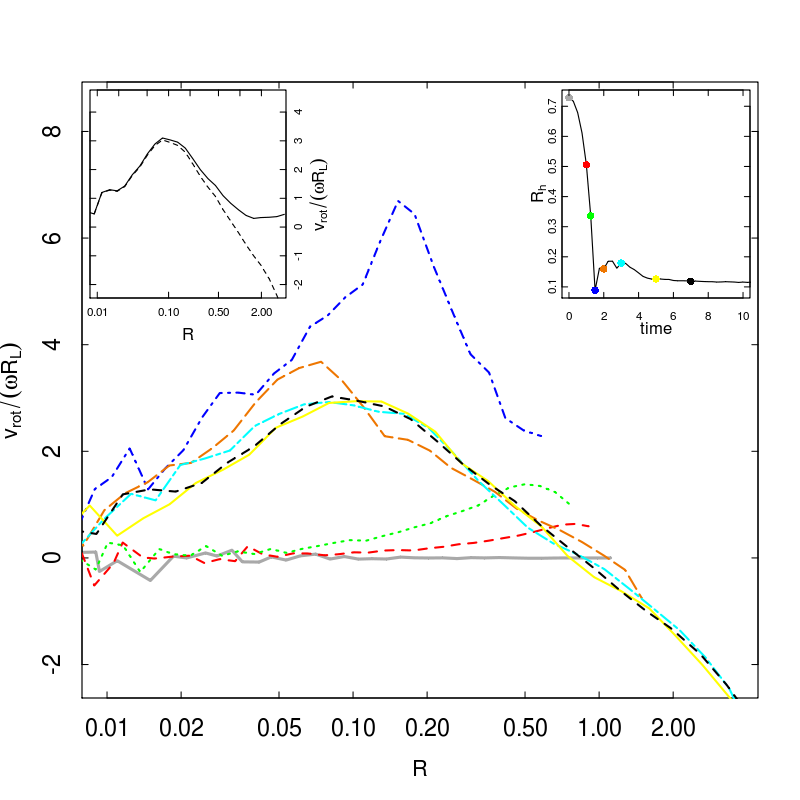}
}
\caption{Radial profile of  the rotational velocity, $V_{rot}$ (normalized to the $\omega R_L$ where $R_L$ is the cluster initial limiting radius and $\omega$ is the cluster angular velocity around the Galactic center),  as
  a function of the cylindrical radius, $R$,  for the \CD simulation at
$t=0$ (grey solid line)
$t=1.0$ (red dashed line)
$t=1.25 $ (green dotted line)
$t=1.5 $ (blue dot-dashed line)
$t=2.0 $ (orange long-dashed line)
$t=3.0 $ (cyan long-short dashed line)
$t=5.0 $ (yellow solid line)
$t=7.0 $ (black dashed line).
The radial profiles are calculated by dividing the system
in cylindrical shells parallel to the $z$ axis. The top left inset shows the rotation curve at $t=7.0$, as measured in the non-rotating (solid line) and rotating (dashed line) coordinate systems. The top right inset shows the time evolution of the 2D half-mass radius with dots showing the times at which the radial profiles of $V_{rot}$ in the main panel are calculated (the color scheme is the same as in the main panel). All the quantities presented in the panels are in N-body units.
 }
\label{fig:vrot}
\end{figure}

The effect of the Coriolis terms in the equation of motion   is illustrated in 
Fig. \ref{fig:vrot}.
This figure
shows the time evolution of the  radial profile of the
rotational velocity around the $z$ axis for  the \CD simulation and illustrates how the system acquires  an internal differential rotational motion during the violent relaxation phase.

During the initial
collapse, the whole system acquires a prograde differential
rotation. After the maximum collapse phase, part of the system rapidly
settles to form an inner high-density core in prograde
rotation. 

For shells of stars re-expanding and moving outward the Coriolis force
acts in the direction opposite to that in which it acted
during the collapse phase; this leads to a decrease in the rotational velocity of stars in the outer regions, with the
outermost shells acquiring a significant
retrograde rotation. The final rotational velocity
profile is characterized by an inner rising portion, a peak in the
cluster intermediate regions followed by  a declining portion in the
cluster outermost regions. We point out that, as already stated
in \S 2, the rotational velocity profiles shown in Fig. \ref{fig:vrot} are
calculated in the coordinate system  synchronously rotating with the
cluster orbital motion around the host galaxy. These profiles therefore show
the development of the cluster differential  rotation relative to the synchronous (angular
velocity equal to the orbital $\omega$) solid-body rotation. In the top left inset in Fig. \ref{fig:vrot} we also show one of the rotational
velocity profiles calculated in the non-rotating coordinate system centered on the cluster center. Since the angular velocity of the cluster central regions
is much higher than the angular velocity of the synchronous rotation $\omega$, the inner rotational velocity profile is barely affected by this change of coordinates. In the outer regions the magnitude of the retrograde rotation in the rotating coordinate system is comparable to that of the synchronous solid-body rotation and, in the non-rotating system, the
rotational velocity tends to small positive values (smaller than the
synchronous rotational velocity). 
The rotational velocity profiles in the two coordinate systems both
illustrate how, as a result of  the collapse and the
subsequent re-expansion during the violent relaxation phase the
cluster acquires  a differential rotational velocity
significantly more (less) rapid  than the initial synchronous rotation in its
inner (outer) regions. 
 \begin{figure}    
\centering{
\includegraphics[width=8.8cm]{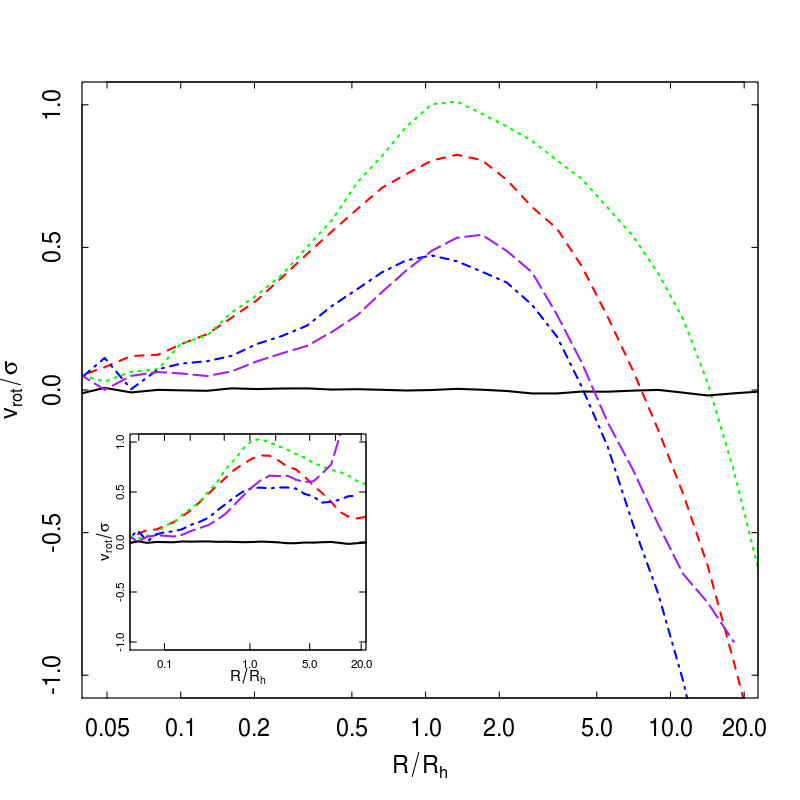}
}
\caption{Radial profiles of the ratio of the rotational velocity around the $z$ axis to the 1-D velocity dispersion, $V_{rot}/\sigma$, as a function of the ratio of the cylindrical radius to the projected half-mass radius, $R/R_h$, for the 
\CD (red dashed line), 
\CH (green dotted line), 
\WD (blue dot-dashed line), 
\CCH (purple long-dashed line), 
and \CI (black solid line) 
simulations. 
The radial profiles shown  are calculated by averaging the profiles at $5\ltorder t/t_{ff}\ltorder 6.5$ (for the \CCH system the profile was calculated by averaging profiles  at $12\ltorder t/t_{ff} \ltorder 13.5$ when there was no residual clumpiness in the cluster structure). The inset shows the profiles calculated in the non-rotating coordinate system.}
\label{fig:vsigma}
\end{figure}

In Fig.\ref{fig:vsigma} we plot the radial profile of the ratio of the rotational
velocity to the 1-D velocity dispersion, $V_{rot}/\sigma$, for all the
simulations (the 1-D velocity dispersion is defined  here as $\sigma=\sqrt{(\sigma_r^2+\sigma_{\theta}^2+\sigma_{\phi}^2)/3}$ as a function of the cylindrical radius $R$). 
All the  systems discussed here are characterized by a
similar shape of the radial $\vsigma$ profile. As was to be expected, the less deep collapse of the cluster starting with a larger initial virial ratio (the \WD simulation)
leads to a final system characterized by smaller values of
$\vsigma$. For all the systems the peak of $\vsigma$ is located at a
distance from the cluster center between $R_h$ and $2R_h$ (where $R_h$
is the cluster projected half-mass radius). The radial profile of $\vsigma$
for the isolated system (\CI) is also shown in this figure and does not show any rotation.   
For completeness, as done in Fig. \ref{fig:vrot}, we show  in the inset the radial profiles of $\vsigma$ calculated in the non-rotating coordinate system.

\begin{figure}    
\centering{
\includegraphics[width=8.8cm]{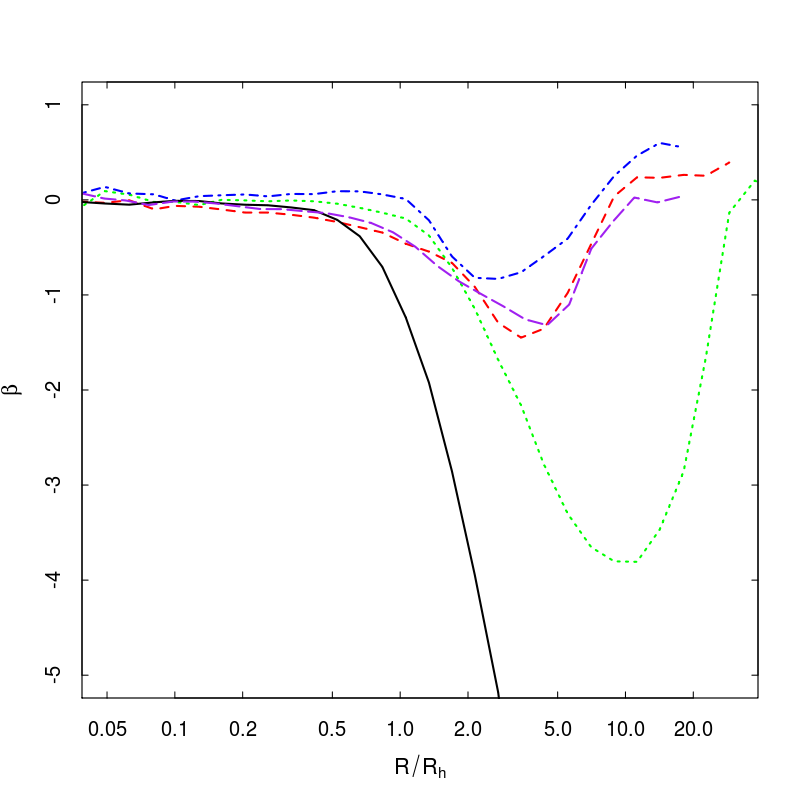}
}
\caption{
Radial profile of the anisotropy parameter $\beta$ as a function of the ratio of the cyclindrical radius to the projected half-mass radius, $R/R_h$, for the 
\CD (red dashed line), 
\CH (green dotted line), 
\WD (blue dot-dashed line), 
\CCH (purple long-dashed line), 
and \CI (black solid line) simulations.
The radial profiles shown  are calculated by averaging the profiles at
$5\ltorder t/t_{ff}\ltorder 6.5$ (for the \CCH system the profile was
calculated by averaging profiles  at $12\ltorder t/t_{ff} \ltorder
13.5$ when there was no residual clumpiness in the cluster
structure).}
\label{fig:beta}
\end{figure}

Finally in Fig.\ref{fig:beta} we plot for all the simulations the radial profiles of
the anisotropy parameter
$\beta=1-2\sigma_r^2/(\sigma_{\theta}^2+\sigma_{\phi}^2)$. For our \CI
simulation we recover the typical radial profile found at the end of
violent relaxation of isolated systems; the system 
is characterized by  
an inner isotropic core and  an increasingly radially anisotropic
outer halo.

 The systems evolving in an external tidal field are characterized by
 a qualitatively similar radial variation of $\beta$ in the inner
 regions but the radial profiles of $\beta$ for these systems 
 deviate from that of an isolated system in the cluster outer
 regions. Specifically, for systems evolving in an external tidal
 field, $\beta$ reaches a minimum  and then rises again until the
 outermost shells where the system becomes isotropic or tangentially
 anisotropic. For the systems we have explored the position of the
 minimum in $\beta$ is at a radius larger than $3R_h$; in units of the
 Jacobi radius the minimum of $\beta$ is in all cases between $0.2R_J$ and $0.4R_J$.  

It is interesting to notice that the behavior of  the anisotropy parameter
emerging from our simulations is qualitatively similar to the one
observed  in $\omega$ Cen (see van der Marel \& Anderson 2010, van de Ven
et al. 2006, and van Leeuwen et al. 2000). This observed kinematical
property is particularly relevant to our study since $\omega$ Cen is
characterized by a very
long half-mass relaxation time and it therefore might be difficult to
ascribe such a feature to the effects of long-term dynamical
evolution (see e.g. Takahashi \& Lee 2000, Baumgardt \& Makino 2003, Hurley \& Shara 2012 for some studies showing that tangential 
anisotropy in the outermost regions of star clusters can arise during
their long-term evolution driven by two-body relaxation; in those
cases the tangential anisotropy has been ascribed to the preferential
loss of stars on radial orbits). 

On the theoretical side, such a behaviour of the anisotropy
parameter is also consistent  with the kinematical properties  of a
family of self-consistent models with differential rotation presented
by Varri \& Bertin (2012) and recently applied to the dynamical
interpretation of selected rotating Galactic globular clusters (Bianchini et al. 2013).

Finally we wish to emphasize that while the early collapse
associated with the violent relaxation phase  is necessary to produce the
rapid inner rotation, the slow outer differential rotation
found in our simulations could also arise during other dynamical
phases driven by processes that cause the clusters to expand
radially (for example, the radial expansion induced by mass
loss due to stellar evolution).

\section{Conclusions}

We have studied the effects of the host galaxy tidal field on the dynamics of a star cluster  violent relaxation.
We have carried out a number of N-body simulations starting
from initial conditions characterized by different density profiles,
tidal fields, and different
initial virial ratios.

We focused our attention on the kinematical properties of the
equilibrium systems emerging at the end of the violent relaxation
phase and we have shown that they differ significantly from those
of systems undergoing this evolutionary phase in isolation. 
Specifically:
1) All the systems we have investigated acquire  a significant
differential rotation around the $z$ axis during the violent
relaxation phase and are characterized by final peak values of $V_{rot}/\sigma$ between $0.5$ and $1$.  
Relatively to a reference coordinate system characterized by
synchronous solid-body rotation (i.e. solid-body rotation with angular velocity equal to that of the cluster orbital motion around the host galaxy),   
the inner regions of the final equilibrium system rotate in the prograde
direction (i.e. their rotational velocity is faster than the synchronous solid-body rotation) while  stars in the outermost regions stars rotate in the
opposite direction (i.e. their rotational velocity is slower than the synchronous solid-body rotation). The typical rotation velocity profile
at the end of our simulations rises from the
cluster center and reaches a maximum at a radius $R_{peak} \sim (1-2)R_h$.
2) The rapid inner rotation is acquired by the system
during the collapse, while the subsequent re-expansion of the outer cluster
shells is responsible for the outer slow rotation.    
While the initial collapse, typical of the first stages of evolution
of a subvirial stellar system, is necessary to produce the inner
rapid rotation, the outer slow differential rotation (slower than the synchronous
solid-body rotation and, therefore, retrograde if measured relatively to
the synchronously rotating
coordinate system) can arise during other evolutionary
phases driven by processes causing a radial expansion of the
cluster (for example during the cluster expansion triggered by mass loss due to
massive star evolution).
3) The presence of an external tidal field affects the radial profile
of the velocity anisotropy. Similarly to clusters evolving
in isolation, the systems explored in this Letter are characterized by
an inner isotropic core, followed by a region of increasing radial
anisotropy. However, for systems evolving in an external tidal field,
the anisotropy parameter
$\beta=1-2\sigma_r^2/(\sigma_{\theta}^2+\sigma_{\phi}^2)$ reaches a
minimum (corresponding to a maximum in the radial anisotropy) and
rises again in the outer 
regions with the outermost regions of the clusters characterized by isotropy or mild tangential anisotropy.   

In this Letter, we have shown that significant differential rotation and a distinct radial variation of the anisotropy can
emerge during a cluster  early evolution.
An extensive study of the dependence of the final equilibrium properties on the cluster initial density and rotational velocity profiles, virial ratio, concentration, and stellar mass spectrum is currently in progress and will be presented in a separate paper.
To establish a connection between the kinematical properties
identified in this study and those of dynamically old clusters, we are
also currently exploring how these kinematical properties change during the
cluster long-term evolution driven by two-body relaxation. 
Young or old but dynamically young clusters, on the other hand, might still keep memory
of the kinematical fingerprints of this early evolution.  
High-precision proper motion studies of star clusters by means of HST and GAIA (see e.g. Bellini
et al. 2013, Pancino et al. 2013) and ESO/VLT radial velocity studies (see e.g. Lanzoni et al. 2013) will help to shed
light on the 
connection between their kinematical properties and their early and
long-term dynamical history.  

{}

\end{document}